# Evaluation of Google's Voice Recognition and Sentence Classification for Health Care Applications

Md Majbah Uddin[1], Nathan Huynh[2*], Jose M. Vidal[3], Kevin M. Taaffe[4],
Lawrence D. Fredendall[5] and Joel Greenstein[6]

*[1]Department of Civil and Environmental Engineering, University of South Carolina*
*[2] Department of Civil and Environmental Engineering, University of South Carolina*
*[3] Department of Computer Science and Engineering, University of South Carolina*
*[4&6]Department of Industrial Engineering, Clemson University*
*[5]Department of Management, Clemson University*

*Emails: [1]muddin@email.sc.edu, [2]nathan.huynh@sc.edu, [3]vidal@sc.edu, [4]taaffe@clemson.edu,*
*[5]flawren@clemson.edu, [6]iejsg@clemson.edu*

<u>*Corresponding author</u>
Civil & Environmental Engineering
College of Engineering and Computing
University of South Carolina
300 Main Street, Columbia, SC 29208
Phone: (803) 777-8947
E-mail: <u>nathan.huynh@sc.edu</u>

**Abstract**

This study examined the use of voice recognition technology in perioperative services (Periop) to enable Periop staff to record workflow milestones using mobile technology. The use of mobile technology to improve patient flow and quality of care could be facilitated if such voice recognition technology could be made robust. The goal of this experiment was to allow the Periop staff to provide care without being interrupted with data entry and querying tasks. However, the results are generalizable to other situations where an engineering manager attempts to improve communication performance using mobile technology. This study enhanced Google's voice recognition capability by using post-processing classifiers, i.e., bag-of-sentences, support vector machine, maximum entropy. The experiments investigated three factors (original phrasing, reduced phrasing, and personalized phrasing) at three levels (zero training repetition, 5 training repetitions, and 10 training repetitions). Results indicated that personal phrasing yielded the highest correctness and that training the device to recognize an individual's voice improved correctness as well. Although simplistic, the bag-of-sentences classifier significantly improved voice recognition correctness. The classification efficiency of the maximum entropy and support vector machine algorithms was found to be nearly identical. These results suggest that engineering managers could significantly enhance Google's voice recognition technology by using post-processing techniques, which would facilitate its use in health care and other applications.

**Introduction**

All engineering managers are faced with the task of efficiently and effectively coordinating multiple workflows to accomplish a task. Many engineering managers use information technology to facilitate communication and reduce coordination costs between sub-tasks. Currently, mobile technologies are being integrated into knowledge work to simplify communication and coordination. For example, mobile technology has simplified the creation of blogs and other creative works (Greenwald, 2014).

Many work processes that require extensive real-time communication can potentially benefit from reliable hands-free use of mobile technology. But, hands-free communication requires an accurate voice recognition system (VRS). Improved VRSs could potentially reduce ineffective communication, a common barrier to system improvement. For example, Adams and Ruiz-Ulloa (2003) found that attempts to install and use Kanbans were hindered by poor communication, while Costa et al. (2013) found that information and communication technology significantly impacted problems in new product development.

The research setting for this study is a Perioperative Services (Periop) unit, a complex environment requiring frequent real-time communication between various staff members located throughout the unit. Periop is a demanding work environment in which multiple, expensive resources must be flexibly coordinated. This research tests the use of VRS in mobile computing technology for rapid, on-going communication between key personnel about the status of each patient. This study examines how to improve VRS in mobile computing technology so that communication frequency between staff could be increased by making the communication hands-free. This in turn should improve coordination between these staff. Specifically, this paper investigated the effectiveness of different post-process algorithms to improve the performance of Google's speech recognizer. Our tests showed that Google's speech recognition system tended to make the same mistakes repeatedly; that is, it returned the same wrong word for a particular sound sample multiple times. To improve the VRS technology, we tested three different algorithms—"bag-of-sentences", support vector machine, and maximum entropy. Although this research tested VRS only in a Periop unit, the findings should be useful to engineering managers in other industries, who may have an interest in using hands-free mobile technology to improve workflow within a unit and across units.

**Literature Review**

Engineering managers have long been concerned with various aspects of improving performance in hospitals and other health care institutions by improving communication. Mazur et al. (2012) examined learning during a lean implementation in a hospital. Gogan et al. (2011) examined how to improve relationships to create a prescription history system that would provide a hospital's emergency department with an accurate prescription history for a patient. Smith et al. (2012) investigated how data standardization could improve performance in the health care supply chain. Twietmeyer et al. (2008) argued that communication is the most basic aspect of knowledge management in any organization, including hospitals.

Voice recognition is the process of creating texts from speech or voice using software. The system records the speech signal, processes the signal and compares the analyzed speech patterns with a collection of possible words and finally, generates the written text (O'Shaughnessy, 2003). Speech recognition technology is not a new concept, though the use of mobile devices using speech recognition is increasing day-by-day. Today's systems have the flexibility to be used in both user dependent and independent domains. User independent systems can be employed by all users without the need to train the system for each individual user, while user dependent systems require training for individual speech patterns (Durling & Lumsden, 2008). Voice recognition technology has matured and advanced significantly in recent years and its potential for health care applications is growing (Zhao, 2009).

Advances in computing power allow current systems to process a large amount of speech data, so that speech recognition technology now has a high of level of accuracy (Zhao, 2009). Moreover, voice recognition has a natural place in the next generation of "smart" environments and has great potential for widespread application (Pentland & Choudhury, 2000). However, there remain challenges, including different speech styles, speech rates, and voice characteristics. (Furui, 2005).

Speech recognition technology could potentially simplify many management tasks. For example, health care generates a large amount of text and documentation, which needs to be accessed quickly (Al-Aynati & Chorneyko, 2003). Health care's traditional documentation method, handwritten records, is time consuming, and dictated records have the added expense of transcription services. Voice recognition is free from these problems as it can immediately transfer spoken words into text (Korn, 1998). Using a voice recognition system, the physician can dictate, edit and create electronic reports instantly; these reports can be made available to other physicians immediately and can be added to the patient record. As a result, the total patient care process can take less time and may lead to better service at a lower cost.

*Current Applications of VRS in the Health Care Setting*

Voice recognition is already being applied in some health care settings. A computer-automated telephone system, known as an Interactive Voice Response System (IVRS), responds when a patient dials a number and selects from a menu of options by pressing the appropriate numbers on the telephone keypad.  The IVRS system leads the patient to a computer network system, which records and documents the voice of the patient and allows the patient to converse with a talking computer.  This interaction includes reminders to refill medication, schedule a clinic visit, check blood pressure, take medication, etc.  The IVRS is an effective data management and reporting system.  However, a common issue is that the system often drops patients during a call. Nonetheless, IVRSs can be a very handy tool for health care services because IVRSs provide live communication (Lee et al., 2003).

The Vocera communication system uses a wearable badge device, which offers a push-to-call button, a small text message screen, and versatile voice-dialing capabilities based on speech recognition.  It also offers hands-free conversation, such as hands-free call and voice message when the recipient is unavailable.  In an experiment in St. Vincent's Hospital, Birmingham, Alabama, the utility of this system was verified.  Another advantage of this system is biometric security, as only the proper user can initiate the call.  The Vocera system can also dial by role or team according to the account information stored on the server (Stanford, 2003).

Alapetite (2008) found that the traditional touch-screen and keyboard interface imposed a steadily increasing mental workload (in terms of items to keep in memory).  In contrast, a speech input interface allowed anesthesiologists to enter medications and observations almost simultaneously.  During time-constrained situations, speech input reduced mental workload related to the memorization of events to be registered because it imposed shorter delays between event occurrence and event registration.  However, existing speech recognition technology and speech interfaces require training to be used successfully.

Voice recognition decreased report turnaround time compared to conventional dictation. However, it performed better when English was the user's first language (Bhan et al., 2008; Mehta et al., 1998; Akhtar et al., 2011). Another viewpoint is that improvement in report turnaround time is correlated with work habits rather than workload (Krishnaraj et al., 2010). Furthermore, radiology reports prepared using VRS had significantly more errors than other methods. Typically, increased errors occurred in noisy areas with high workload and with radiologists whose first language was not English (McGurk et al., 2008).

Rana et al. (2005) found that for long reports voice recognition was advantageous over traditional tape dictation-transcription in total reporting time. Voice recognition methods incorporate dictation and transcription into one stage, whereas dictation-transcription method requires several stages and individuals in the process. Several issues with voice recognition in the radiology department included: (1) inadequate training, (2) insufficient attention to operational issues, (3) an increase in the dictation cost, and (4) an increase in the workload of the radiologist.

Voice recognition has been used in many other hospital departments. Computer-based transcription is a relatively inexpensive alternative to traditional human transcription in pathology where numerous reports must be regularly transcribed (Al-Aynati & Chorneyko, 2003). Voice recognition technology improved the efficiency of workflow, minimized transcription delays and costs, and contributed to improved turnaround time in surgical pathology (Henricks et al., 2002). Emergency departments have used voice recognition systems as a tool for physician charting and have been found to be nearly as accurate as traditional transcription, with shorter turnaround times and lower costs (Zick & Olsen, 2001). Voice recognition technology has been used for nurse dictation (Carter-Wesley, 2009) and has improved workflow in many clinical processes.  However, Issenman and Jaffer (2004) found that computer dictation and correction time was greater using voice recognition than using electronic signatures for letters typed by an experienced transcriptionist in a pediatric gastroenterology unit.

Nuance's Dragon NaturallySpeaking is used with the Apple iPhone.  Parente et al. (2004) found this technology to be very cost effective and acceptable to physicians for filling out different types of forms, as well as in creating an electronic health record (EHR).  Dragon NaturallySpeaking has been used by radiologists to create reports, significantly reducing turnaround times and decreasing transcription costs (Donnelly, 2013).

### *Limitations of VRS*

Currently, there are multiple problems with voice recognition software.  Devine et al. (2000) found that Dragon Systems NaturallySpeaking Medical Suite, version 3.0 had the highest error rate among three commercially available continuous speech recognition software packages: (1) IBM ViaVoice 98, (2) Dragon Systems NaturallySpeaking Medical Suite, and (3) L&H Voice Xpress for Medicine.  Murchie and Kenny (1988) found that voice recognition resulted in significantly more errors than keyboard entry.  Moreover, Grasso (1995) found that a speech recognition system had some limitations in terms of vocabulary size, continuity of speech and speaker dependency.  The system needed *a priori* training to verify the capability of the device to act on various conditions.  When the vocabulary size became bigger it needed more time for training.  It could not distinguish multiple word boundaries—as in "youth in Asia" and "euthanasia".  Increasing the size of the vocabulary also adversely affected the accuracy of the system.

## Methodology

### Data Collection Device

For this study, a smart-app named Perioperative Services Mobile Learning System (POS-MLS) was developed by the research team using the latest Android API (Level 19). One portion of this app includes a screen with 16 Pre-op checklist items that could be marked complete using touch or voice. The voice recognition is enabled by the Android platform with its built-in speech recognizer. The Android speech recognizer gathers a sound sample from the user and sends it to Google's cloud-based speech recognition service, which then returns a plain text reply, as a string. The speech recognizer performs a best effort to find the most likely set of words to match the sound sample. We set the language to U.S. English, indicating to the recognizer our choice of spoken language for testing. The data collected for this paper were based on version 0.7 of the smart-app. Exhibit 1 shows a screenshot of the checklist items.

**"Insert Exhibit 1 here."**

### Data Collection Procedure

The smart-app was installed on Google Nexus 4, 7, and 10 mobile devices for the experiments. The experiments investigated three factors, with each factor having three levels. The three factors were: as-is phrase (from the Pre-op checklist items), reduced phrase (developed by the research team), and personalized phrase (selected by the individual participant) [see Exhibit 2]. Each factor had three levels in the experiment: Google-only (zero training repetition), Train-5 (5 training repetitions), and Train-10 (10 training repetitions). In the Google-only case, the app is not 'learning' from prior data. When training is allowed in the Train-5 and Train-10 levels, the app can learn from past mistakes and recognize phrases based on those mistakes. The results collected from the experiments were classified as either correct or incorrect in terms of recognition of the spoken phrase. Various text classification algorithms, used in this study, are described next.

**"Insert Exhibit 2 here."**

### Text Classification Algorithms

A variety of supervised learning algorithms have been using for text classification: naïve Bayes (Lewis, 1998), support vector machine (Dumais et al., 1998), maximum entropy (Nigam et al., 1999) and k-nearest neighbor (Yang, 1999). For this study, we investigated support vector machine (SVM) and maximum entropy (MAXENT), in addition to the simple "bag-of-sentences" approach. A comparison between SVM and MAXENT classifiers can be found in the work by Du and Wang (2012). The rationale for selecting these three algorithms, along with a description of each algorithm, is provided next. The simplest of these three algorithms, "bag-of-sentences", is described first.

### Bag-of-Sentences

During a training round we matched each of the returned phrases to the desired phrase. For example, if we said "administer medications" but the speech recognizer returned "Minister medications" we then added the fact that "Minister medications" should always match "administer medications" to the learning table. If some other spoken phrase returned "Minister medications" then that phrase would always be matched. That is, new matches overwrote old matches during the training phase. In this manner, we created a many-to-few mapping between phrases returned by the speech recognizer and phrases we needed to recognize. This simple bag-of-sentences strategy works well for this application because there are only a few phrases that need to be recognized. Once the training was done, the app uses the table to translate text phrases returned by the speech recognizer into one of the target phrases.

*Support Vector Machine*

Support Vector Machine (Cortes & Vapnik, 1995), a supervised machine learning technique, is gaining much attention due to its superior data classification and regression performance (Pham et al., 2011). SVM has been applied to many fields related to health care for classification problems (Maglogiannis & Zafiropoulos, 2004; Yu et al., 2010; Zhang et al., 2011). The SVM algorithm is based on training, testing and performance evaluation because it is a learning machine. In training, a convex cost function is optimized. In testing the model is evaluated using support vectors to classify a test data set, and performance evaluation is based on error rate determination.

For this text classification study an $\varepsilon$-SVM was adopted—similar to Pham et al. (2011). A text classification problem with $N$ inputs $\{x(i)\}_{i=1}^{N}$, $x(i) \in R^{In}$ and outputs $\{y(i)\}_{i=1}^{N}$, $y(i) \in R^{1}$ is assumed. Using a function $\Phi(x(i))$, the $\varepsilon$-SVM model maps the inputs from the Infinite-dimensional space into a higher $h$-dimensional space. The estimation function of output $y(i)$ has the form specified in Equation (1).

$$\hat{y}(i) = f(x(i)) = w^{T}\Phi(x(i)) + b \tag{1}$$

The coefficients, $b \in R^{1}$ and $w \in R^{h}$, are estimated using Equations (2) – (5).

Minimize $\qquad \dfrac{1}{2}w^{T}w + \dfrac{C}{N}\sum_{i=1}^{N}\left(\xi_{i} + \xi_{i}^{*}\right) \tag{2}$

Subject to
$$w^T \Phi(x(i)) + b - y(i) \le \varepsilon + \xi_i \qquad (3)$$

$$y(i) - w^T \Phi(x(i)) - b \le \varepsilon + \xi_i^* \qquad (4)$$

$$\xi_i, \xi_i^* \ge 0, \qquad i = 1, \ldots, N \qquad (5)$$

Here $\xi_i$ and $\xi_i^*$ = slack variables, C = a regularization parameter, T = transpose, and $\varepsilon$ = soft margin loss parameter. If the difference between $\hat{y}(i)$ and $y(i)$ is larger than $\varepsilon$, $\xi_i$ or $\xi_i^*$ can only be greater than zero (Exhibit 3).

**"Insert Exhibit 3 here."**

*Maximum Entropy*

Maximum entropy classification has been shown to be an effective alternative technique in a number of natural language processing applications (Berger et al., 1996). Its application for text classification was proposed by Nigam et al. (1999). The following provides a brief review of the maximum entropy algorithm and explains how it classifies texts (refer to Nigam et al. (1999) for additional details).

Training data is used to set constraints on the conditional distribution. When any real-valued function of the document and class is a feature, $f_i(d,c)$, the model distribution will have the same expected value for this feature similar to the training data, $D$. Then, the learned conditional distribution, $P(c \mid d)$, must have the property specified in Equation (6).

$$\frac{1}{D} \sum_{d \in D} f_i(d, c(d)) = \sum_d P(d) \sum_c P(c \mid d) f_i(d, c) \qquad (6)$$

And, the distribution of $P(c \mid d)$ has an exponential form (Della Pietra et al., 1997), where each $f_i(d,c)$ is a feature/class function for feature $f_i$, $Z(d)$ is a normalization factor to ensure proper probability and $\lambda_i$ is a parameter, as specified in Equation (7).

$$P(c \mid d) = \frac{1}{Z(d)} \exp\left( \sum_i \lambda_i f_i(d,c) \right) \tag{7}$$

Word counting is a feature of text classification with maximum entropy, since applying maximum entropy to a domain requires the selection of a set of features to use for setting the constraints. For each word-class combination the feature is considered, as shown in Equation (8).

$$f_{w,c'}(d,c) = \begin{cases} 0 & \text{if } c \neq c' \\ \dfrac{N(d,w)}{N(d)} & \text{Otherwise} \end{cases} \tag{8}$$

where $N(d,w)$ is the number of times word $w$ occurs in document $d$, and $N(d)$ is the number of words in $d$.

It is expected that features accounting for the number of times a word occurs should improve classification in text. This implies that the weight for the word-class pair would be higher than for the word paired with other classes if a word occurs often in one class. The tool used to implement these algorithms, RTextTools, is described next.

*RTextTools*

RTextTools is a supervised learning package for text classification (Jurka et al., 2013). It provides a comprehensive approach to text classification and also accelerates the classification process. The statistical software R is essential for using this text classification package. The classification process starts with loading data from a CSV, Access or Excel file by calling a function in R. Then a matrix is generated from the data. Then a container object is created that contains all the objects for further analysis. After that, the data are trained by algorithms. Data classification is done next. Finally, the classification is summarized to find the correct classification, which will give the percentage of correct classifications.

RTextTools can work with nine algorithms for training of data. In our study, we used the support vector machine and maximum entropy algorithms to train our data. RTextTools uses support vector machine from the 'e1071' package (Meyer et al., 2012) and maximum entropy from the 'maxent' package (Jurka, 2012) of R. SVM is used to train a support vector machine, and can be used for general regression and classification. MAXENT is used for low-memory, multinomial logistic regression.

**Experimental Set-up**

We conducted 16 experiments that were designed to test the ability of the app to recognize the Pre-op checklist items correctly using voice. The participants were from various age groups, both genders, native and non-native speakers, various ethnic groups, and had different occupations. All of the participants were provided with a Nexus device with the voice-recognition app (version 0.7) installed on it. In the case of as-is phrases, every phrase was spoken five times for all three levels, i.e., Google-only, Train-5, and Train-10. Thus, we have a total of 80 ($16 \times 5$) observations for each phrase at all three levels. In the Train-5 and Train-10 levels for the as-is phrases, we have an additional 5 and 10 training repetitions of phrases, respectively. Data for the reduced and personalized phrases were collected using a similar procedure, with each having 80 observations at all three levels. Exhibit 4 summarizes the phrases, levels, and corresponding post-processing methods. As noted by the check marks, the Google-only level does not involve any training repetition.

**"Insert Exhibit 4 here."**

## Results

### *Correctness by Level*

Exhibit 5 summarizes the app's ability to correctly recognize as-is phrases over 80 observations. On its own (Google-only), the app correctly identified phrases 4% to 86% of the time, with a median of 34%. In the Google-only level, most phrases were not identified correctly. The four phrases identified correctly less than 15% of the time, included words not frequently used in daily life (e.g., RN, H&P, and heparin). At the Train-5 level, recognition correctness increased to approximately 63% (median) but still exhibited a large range (38% to 91%). Two phrases ("need implants" and "implant(s) available") were not recognized any more often at the Train-5 level. Recognition correctness increased with Train-10. More phrases were also correctly identified. The median was 69%. While the range was smaller (44% to 79%), recognition correctness of three phrases ("patient not ready", "RN complete", and "need marking") decreased when Train-10 was used.

Statistically significant differences in recognition correctness between training levels were identified for 11 of the 16 phrases using a Chi-Squared test. Closer examination of these phrases revealed that phrases relying more heavily on medical terminology, such as 'anesthesia', 'heparin', 'RN', and 'H&P,' were recognized less often. The results do suggest that training can improve classification of many phrases. Phrases consisting of commonly used words, e.g.,

"consent obtained", and "need implants," for example, tended to have high classification scores, irrespective of training level.

**"Insert Exhibit 5 here."**

The second factor replaced the as-is 16 phrases with shorter phrases using less medical-based terminology. Results are summarized in Exhibit 6. On its own (Google-only level), the app correctly recognized 53% (median) of the phrases, with a range of 5% - 76%. Seven of the phrases were identified correctly less often than the corresponding as-is phrases. Statistically significant improvements in recognition correctness were found when Train-5 and Train-10 were used for seven phrases. In addition, in reviewing the results presented in Exhibits 5 and 6, for every phrase, when the Google-only approach did not recognize an as-is or reduced phrase at least half the time, both training levels (Train-5 and Train-10) improved recognition correctness.

**"Insert Exhibit 6 here."**

### Correctness by Phrase Type

Exhibit 7 provides a summary of average recognition correctness percentages for all three phrase types, i.e., as-is, reduced, and personalized. All differences in recognition correctness as a function of training are significant ($p < 0.05$), with the exception of the difference between Train-5 and Train-10 for the as-is phrase ($p = 0.129$). This result suggests that training improves recognition correctness. The average recognition correctness for the as-is phrase was 61% when the app was trained with at least five repetitions. For reduced and personalized phrases, average correctness increase between Train-5 and Train-10. In the case of reduced phrases, a similar improvement was observed. Moreover, the correctness percentages, for all three levels, was always greater than that of the as-is phrases (38% vs 47%, 61% vs 63% etc.).  However, these improvements of correctness over as-is phrases is significant only for Google-only level ($p = 0.025$).  For the personalized phrase, the average correctness percentages, for all the three levels, were the highest.  Average correctness for personalized phrases also increased with training levels.  It is clear that training repetitions improve the app performance, and increasing the number of training repetitions from 5 to 10 continued to increase recognition correctness with the exception of as-is phrases.  In addition, personalized phrases are identified correctly more frequently than as-is and reduced phrases for pre-op checklist items within a voice recognition application, suggesting that personalized phrases may be more suitable.

**"Insert Exhibit 7 here."**

### Correctness by Classifier

For classification using supervised algorithms, training data is required to classify the text.  For that reason we do not have correctness values for the Google-only level.  Exhibit 8 compares the average correctness percentages between the support vector machine (SVM) algorithm and the maximum entropy (MAXENT) algorithm.  It is clear from Exhibits 7(a) and 8 that classification using SVM and MAXENT algorithms improved classification correctness significantly more than the bag-of-sentences approach in most cases (5 out of 6).  Train-5 with as-is phrases yields the maximum average correctness for SVM of 82% and for MAXENT of 84%.  However, those values for Train-10 are within 1% of the Train-5 value.  Unlike the bag-of-sentences approach, increasing training repetitions does not lead to further correctness of classification.  Average correctness results using the reduced phrases show the same decreasing pattern.  Average correctness of SVM decreases from 79% to 77% and MAXENT from 80% to 79% for reduced phrase.  The average correctness for Train-10 is less than Train-5 for both algorithms.  For the personalized phrases, the average correctness value for SVM with Train-10 (77%) was not significantly different than the bag-of-sentences (79%).  The bag-of- sentences was also not significantly different for Train-10 with MAXENT (81%).  Additionally, in case of personalized phrase, $p$-values suggest that with higher levels of training there is no difference between SVM and MAXENT.  The biggest differences in average correctness occurred between bag-of-sentences and supervised algorithms and were 21% for SVM and 23% for MAXENT.  The MAXENT algorithm outperformed SVM for three different cases (as-is, using both Train-5 and Train-10, and personalized, using train-5, only).  There was no difference between SVM and MAXENT for the other three cases.

**"Insert Exhibit 8 here."**

**Conclusions**

This study sought to identify a suitable algorithm to classify phrases in order to improve the performance of Google's speech recognizer to allow hands-free use of mobile technology.  It also examined how system training by the user affected system performance.  Three phrase sets were tested.  The as-is phrases were word-for-word from an existing hospital checklist.  The reduced phrases were developed by the researchers to reduce the number of words and improve word boundaries.  The personalized phrase sets were selected by each, individual user.

As expected the as-is phrases and the Google-only speech recognizer used without any classifier had the lowest phrase recognition correctness in their respective settings.  The use of reduced phrases or personalized phrases improved recognition correctness compared to using the as-is phrase.  The use of two different post-process learning algorithms enhanced speech recognition correctness, compared to the post-process bag-of-sentences approach. Training, i.e., repetitions of phrases, significantly increased speech recognition correctness for all levels of post-processing.

As engineering managers consider the use of voice recognition software to increase coordination within and between work units they need to recognize that to achieve reduced flow times and reduced costs they must adapt existing technology to suit users. These results suggest that it is important to reduce the number of phrases and the number of words in phrases. Another important point for the engineering manager to note is that performance is improved by making word boundaries more distinct. This can be accomplished by providing direction to the users on how to create personalized phrases that have distinct word boundaries. In addition, the manager must ensure that the system allows for post-processing after initial word recognition by the system. Overall, Google's speech recognizer was significantly enhanced by the use of post-processing techniques.

Voice recognition provides two sources of value for perioperative services. First, it provides timelier recording of data, which then allows others to use data for their own processes in real time. Second, it allows for increased data accuracy because data are recorded as work occurs, rather than from memory. By providing timelier and accurate data, VRSs should allow perioperative services managers to develop evidence-based management practices. Changes to the system can be rapidly tested to see if they increase on-time starts and on-time finishes, and whether or not such changes translate into improved surgeon satisfaction, decreased morbidity and/or reduced post-operative wound infections. The potential value of data gathered through VRSs comes with little risk. Delays in recording data and frequent errors have led hospitals to invest in departments that "clean" data to avoid billing errors. As hospitals start to use VRS they may retain data checking initially, to ensure that detectable errors in VRS data are eliminated.

Research to study how to integrate speech recognition engines with artificial intelligence techniques is a promising area for future research. Such research could be used to identify patterns in processes and might allow for using smart-apps to collect additional data. Artificial intelligence could also be used, in conjunction with real time data from smart apps, to identify bottlenecks and suggest interventions to nurse managers before patient flows are affected. An artificial intelligence component would allow a smart-app to learn preferences of each staff member that, in turn, would enable the app to provide meaningful, context-sensitive and personalized information to each staff member. For example, if an anesthesiologist prefers to visit patients in the pre-operative room, followed by visits to the operating rooms, the smart-app could recognize this route preference and send an alert to the anesthesiologist when he/she is deviating from the usual route. Such an alert would essentially create a *poka-yoke* device that eliminates defects resulting from deviations by human operators (Shingo, 1989).

Using VRS with a smart-app has applications for health care that go far beyond its potential ability to improve performance within Perioperative Services. One possible application with great potential is to simplify an individual's ability to gather information on a daily basis and to use these data to predict the likelihood of a particular disease or identify current state of "wellness." Apps are being developed to record multiple types of data that can be used by health care clinicians to help patients improve their health. For example, improved speech recognition may make it easier for patients to record valuable information about their food consumption each day and to share this information with health care providers. VRS can assist greatly with the recording of this information. Artificial intelligence could then be used to analyze this information and provide the clinician with a summary, which can be used to determine if health care intervention is needed.  These are just a few examples of opportunities that exist within this domain.  Additional research, similar to the study elaborated in this article, is needed to help clinicians and engineering managers determine the potential impact of these opportunities.

**Acknowledgement**

This research was supported by the National Science Foundation under grant IIS-SHB #1237007.  The authors would like to thank the two reviewers for their feedback and the editors for their meticulous proofreading of our manuscript.  We truly appreciate their time, effort, and guidance.  Our manuscript has been significantly enhanced by their efforts.

**Exhibit 1.** Screenshot of the Smart-App (POS-MLS)

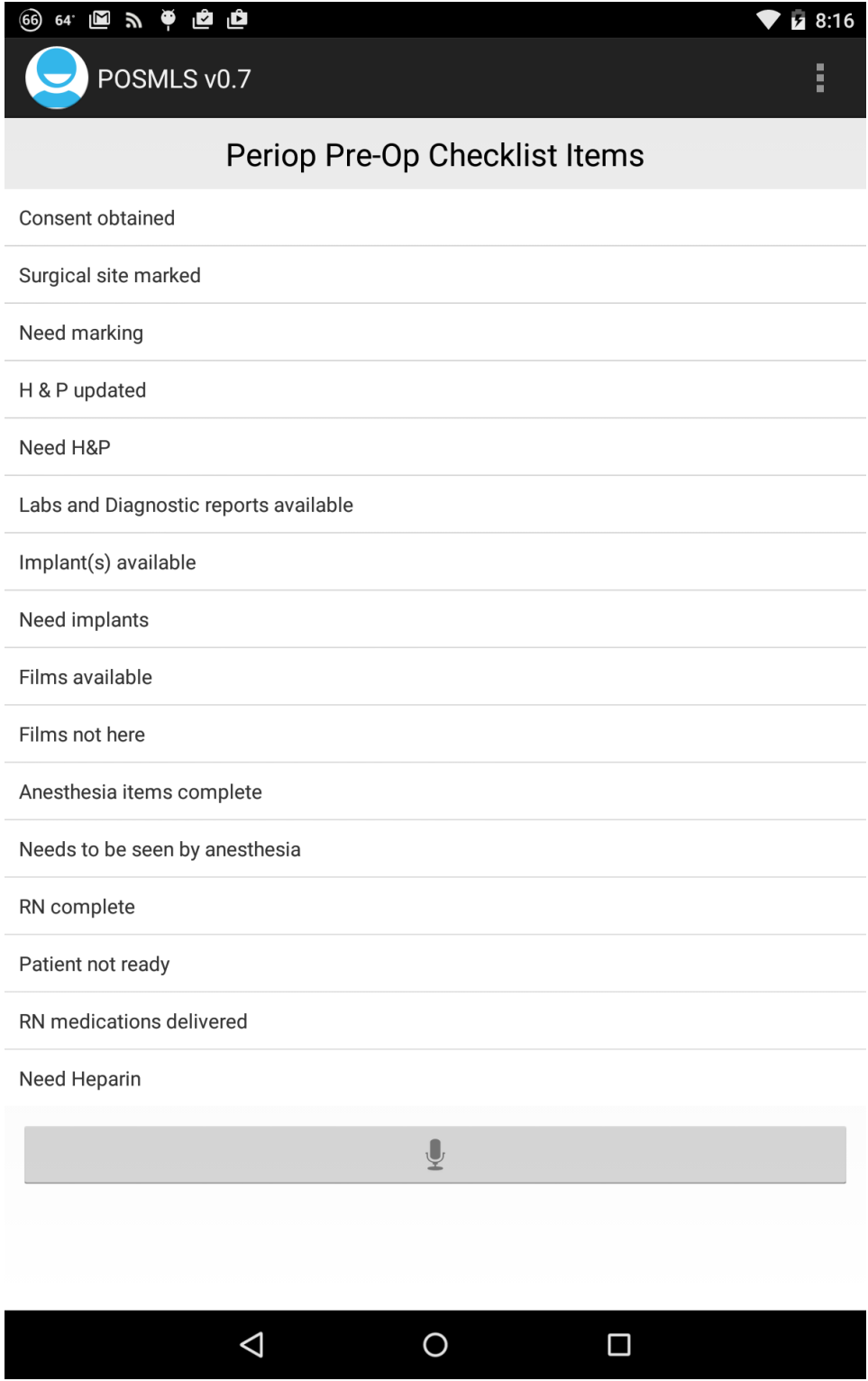

**Exhibit 2**. Types of Phrases

| As-is Phrase | Reduced Phrase | Personalized Phrase* |
|---|---|---|
| Consent obtained | Have consent | Consent good |
| Surgical site marked | Site marked | Site marked |
| Need marking | Need site marked | Need marking |
| H&P updated | History Physical updated | HP good |
| Need H&P | Need History Physical | Need HP |
| Labs and diagnostic reports available | Reports ready | Reports ready |
| Implant(s) available | Implants ready | Implants ready |
| Need implants | Need implants | Get implants |
| Films available | Films here | Films here |
| Films not here | Need films | Need films |
| Anesthesia items complete | Anesthesia complete | Anesthesia done |
| Need to be seen by anesthesia | Need anesthesia | Need anesthesia |
| RN complete | Nurse done | Nurse done |
| Patient not ready | Patient not done | Patient not done |
| RN medications delivered | Medications delivered | Meds given |
| Need heparin | Need heparin | Need hep |

* Each participant created his/her own personalized phrase

**Exhibit 3.** Soft Margin Loss Parameter in ε-SVM (Pham et al., 2011)

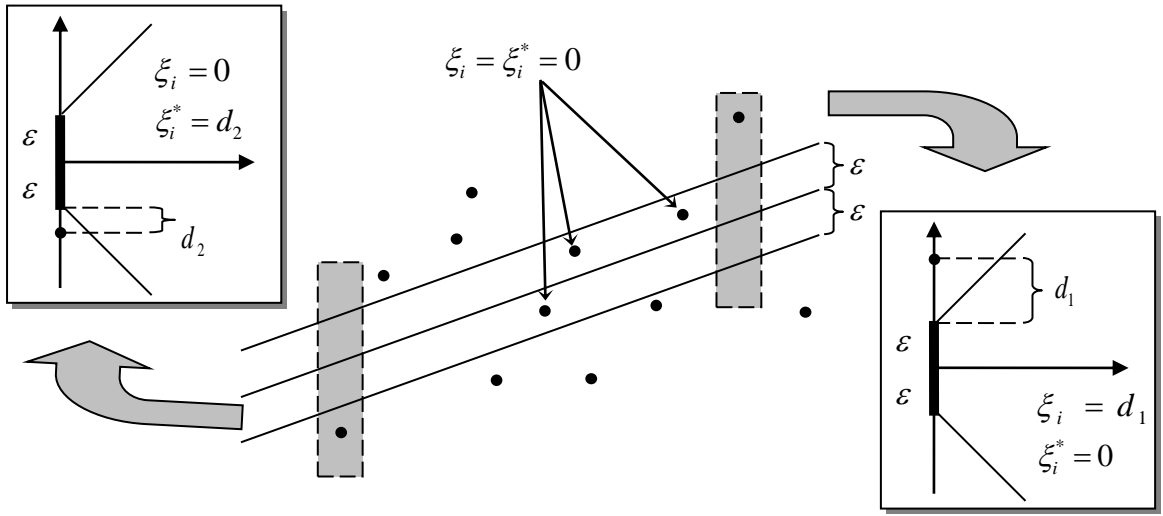

**Exhibit 4.** Summary of Experimental Set-up

| Phrases | Training Repetitions | Testing Repetitions | Google-only | Post-Processing Methods | | |
|---|---|---|---|---|---|---|
| | | | | Bag-of-sentences | Support Vector Machine | Maximum Entropy |
| As-is | 0 | 5 | ✓ | | | |
| | 5 | 5 | | ✓ | ✓ | ✓ |
| | 10 | 5 | | ✓ | ✓ | ✓ |
| Reduced | 0 | 5 | ✓ | | | |
| | 5 | 5 | | ✓ | ✓ | ✓ |
| | 10 | 5 | | ✓ | ✓ | ✓ |
| Personalized | 0 | 5 | ✓ | | | |
| | 5 | 5 | | ✓ | ✓ | ✓ |
| | 10 | 5 | | ✓ | ✓ | ✓ |

**Exhibit 5.** Comparison of Percent Correct and Number of Correct Classifications at Different Training Levels for As-is Phrases

| As-is Phrase | % Correct Classification (Number of Correct Classification) | | | *p*-Value |
| --- | --- | --- | --- | --- |
| | Google-only | Train-5 | Train-10 | |
| Consent obtained | 66.3 (53) | 73.8 (59) | 75.0 (60) | 0.414 |
| Surgical site marked | 28.8 (23) | 53.8 (43) | 57.5 (46) | <0.001 |
| Need marking | 31.3 (25) | 65.0 (52) | 63.8 (51) | <0.001 |
| H&P updated | 40.0 (32) | 62.5 (50) | 70.0 (56) | <0.001 |
| Need H&P | 11.3 (9) | 50.0 (40) | 53.8 (43) | <0.001 |
| Labs and diagnostic reports available | 18.8 (15) | 41.3 (33) | 43.8 (35) | 0.001 |
| Implant(s) available | 65.0 (52) | 65.0 (52) | 75.0 (60) | 0.292 |
| Need implants | 75.0 (60) | 75.0 (60) | 76.3 (61) | 0.978 |
| Films available | 57.5 (46) | 66.3 (53) | 70.0 (56) | 0.237 |
| Films not here | 40.0 (32) | 61.3 (49) | 73.8 (59) | 0.005 |
| Anesthesia items complete | 28.8 (23) | 37.5 (30) | 53.8 (43) | 0.001 |
| Need to be seen by anesthesia | 37.5 (30) | 62.5 (50) | 65.0 (52) | <0.001 |
| RN complete | 8.8 (7) | 81.3 (65) | 76.3 (61) | <0.001 |
| Patient not ready | 86.3 (69) | 91.3 (73) | 78.8 (63) | 0.079 |
| RN medications delivered | 3.8 (3) | 50.0 (40) | 67.5 (54) | <0.001 |
| Need heparin | 5.0 (4) | 46.3 (37) | 60.0 (48) | <0.001 |

**Exhibit 6.** Comparison of Percent Correct and Number of Correct Classifications at Different Training Levels for Reduced Phrases

| Reduced Phrase | % Correct Classification (Number of Correct Classification) | | | *p*-Value |
|---|---|---|---|---|
| | Google-only | Train-5 | Train-10 | |
| Have consent | 63.8 (51) | 72.5 (58) | 77.5 (62) | 0.151 |
| Site marked | 7.5 (6) | 57.5 (46) | 70.0 (56) | <0.001 |
| Need site marked | 12.5 (10) | 36.3 (29) | 52.5 (42) | <0.001 |
| History Physical updated | 50.0 (40) | 47.5 (38) | 53.8 (43) | 0.729 |
| Need History Physical | 37.5 (30) | 41.3 (33) | 47.5 (38) | 0.433 |
| Reports ready | 67.5 (54) | 76.3 (61) | 91.3 (73) | 0.001 |
| Implants ready | 53.3 (43) | 70.0 (56) | 72.5 (58) | 0.026 |
| Need implants | 75.0 (60) | 76.3 (61) | 77.5 (62) | 0.933 |
| Films here | 28.8 (23) | 75.0 (60) | 77.5 (62) | <0.001 |
| Need films | 30.0 (24) | 68.8 (55) | 72.5 (58) | <0.001 |
| Anesthesia complete | 51.3 (41) | 61.3 (49) | 67.5 (54) | 0.107 |
| Need anesthesia | 53.8 (43) | 53.8 (43) | 75.0 (60) | 0.006 |
| Nurse done | 56.3 (45) | 65.0 (52) | 68.8 (55) | 0.242 |
| Patient not done | 76.3 (61) | 75.0 (60) | 77.5 (62) | 0.933 |
| Medications delivered | 75.0 (60) | 83.8 (67) | 83.8 (67) | 0.268 |
| Need heparin | 5.0 (4) | 42.5 (34) | 57.5 (46) | <0.001 |

**Exhibit 7.** Comparison of Average Correctness Percentages for Different Phrase Types

**(a)**

| | Google-only | | Train-5 | | Train-10 | | $p$-Value[a] | $p$-Value[b] | $p$-Value[c] |
|---|---|---|---|---|---|---|---|---|---|
| | Average | Std. dev. | Average | Std. dev. | Average | Std. dev. | | | |
| As-is | 37.7 | 11.2 | 61.4 | 17.9 | 66.3 | 18.7 | <0.001 | <0.001 | 0.129 |
| Reduced | 46.5 | 22.3 | 62.7 | 14.5 | 70.2 | 15.9 | 0.003 | <0.001 | <0.001 |
| Personalized | 53.8 | 22.7 | 72.3 | 16.2 | 78.7 | 12.7 | <0.001 | <0.001 | 0.002 |

[a] Test between Google-only and Train-5.
[b] Test between Google-only and Train-10.
[c] Test between Train-5 and Train-10.

**(b)**

| Test Variable | $p$-Value | | |
|---|---|---|---|
| | Google-only | Train-5 | Train-10 |
| As-is and Reduced | 0.025 | 0.382 | 0.127 |
| As-is and Personalized | <0.001 | 0.007 | 0.006 |
| Reduced and Personalized | 0.022 | <0.001 | 0.003 |

**Exhibit 8.** Comparison of Average Correctness Percentages among the Classifiers

| | SVM | | MAXENT | | $p$-Value[a] | $p$-Value[b] | $p$-Value[c] |
|---|---|---|---|---|---|---|---|
| | Average | Std. dev. | Average | Std. dev. | | | |
| As-is | | | | | | | |
| Train-5 | 81.9 | 11.8 | 84.0 | 9.4 | <0.001 | <0.001 | 0.018 |
| Train-10 | 80.9 | 8.7 | 83.8 | 7.7 | <0.001 | <0.001 | 0.022 |
| Reduced | | | | | | | |
| Train-5 | 78.6 | 14.1 | 80.2 | 9.9 | <0.001 | <0.001 | 0.166 |
| Train-10 | 77.4 | 15.5 | 79.1 | 13.1 | 0.004 | <0.001 | 0.114 |
| Personalized | | | | | | | |
| Train-5 | 79.0 | 13.0 | 81.3 | 13.5 | 0.001 | <0.001 | 0.015 |
| Train-10 | 76.7 | 14.5 | 80.6 | 11.6 | 0.292 | 0.222 | 0.052 |

[a] Test between Bag-of-sentences and SVM.

[b] Test between Bag-of-sentences and MAXENT.

[c] Test between SVM and MAXENT.

**About the Authors**

**Md Majbah Uddin** is a graduate student in civil and environmental engineering at the University of South Carolina. He received a BS in civil engineering from the Bangladesh University of Engineering and Technology. His research interests are in freight transportation planning, routing, text mining, and machine learning.

**Nathan Huynh** is an associate professor at the University of South Carolina. He obtained his Ph.D. and M.S. degrees in Transportation Engineering from the University of Texas at Austin. His research interests include intermodal network design, freight transportation systems, and health care systems. He has published over 50 peer-reviewed papers in these areas. His research has been funded by various federal and state agencies, including NSF, NIH, NIST, and SCDOT.

**Jose M Vidal** is a professor at the University of South Carolina. His research interests include multiagent systems, agent-based modeling and simulations, collaborative technologies, and web applications. He has over ninety publications in major conferences and journals and has received funding from the NSF and DARPA, including an NSF CAREER award. He received a Ph.D. from the University of Michigan and a BSE from the Massachusetts Institute of Technology, both in Computer Science and Engineering.

**Kevin M. Taaffe**, Ph.D., has 25 years of industry and academic experience using simulation and other mathematical modeling tools to solve problems in health care and humanitarian logistics, as well as in transportation and productions systems. His research has been funded by various private and federal entities.  In particular, he addresses health care logistics research ranging from patient flow to operating room scheduling to emergency evacuation.  He serves as a Region Vice President for IIE, as well as a member of INFORMS and the OR Society.

**Lawrence Fredendall** obtained a M.B.A. in 1986 and a Ph.D. in 1991, both from Michigan State University with a supply chain management focus.  Since 1990 he has taught at Clemson University and for the past six years he has investigated the effectiveness of using lean tools in hospitals' perioperative services.  The National Science Foundation (NSF) is funding his research to examine how mobile technology apps can improve resource coordination within perioperative services.

**Joel Greenstein** is an Associate Professor of Industrial Engineering at Clemson University.  His degrees are in mechanical engineering with an emphasis on product and system design from the University of Illinois at Chicago (BS), Stanford University (MS), and the University of Illinois at Urbana-Champaign (Ph.D).  Dr. Greenstein's research interests are in user-centered design methodology and human-computer interaction.  He is a Fellow of the Human Factors and Ergonomics Society.